\documentclass[prl,twocolumn,showpacs,superscriptaddress,groupedaddress]{revtex4-1}
\usepackage{amsmath}
\usepackage{amsfonts}
\usepackage{amssymb}
\usepackage{graphicx}
\usepackage{graphicx,amssymb,amsbsy,color}

\input epsf
\epsfverbosetrue

\begin{document}

\title{Fresnel aperture diffraction: a phase-sensitive probe for
superconducting pairing symmetry}
\author{C. S. Liu$^{1,2}$ and W. C. Wu$^1$}
\affiliation{$^1$Department of Physics, National Taiwan Normal University, Taipei 11677,
Taiwan\\
$^2$Department of Physics, Yanshan University, Qinhuangdao 066004, China}

\begin{abstract}
Fresnel single aperture diffraction (FSAD) is proposed as a phase-sensitive
probe for pairing symmetry and Fermi surface of a superconductor. We
consider electrons injected, through a small aperture, into a thin
superconducting (SC) layer.
It is shown that in case of SC gap symmetry $\Delta(-k_x,\mathbf{k}%
_\parallel)=\Delta(k_x,\mathbf{k}_\parallel)$ with $k_x$ and $\mathbf{k}%
_\parallel$ respectively the normal and parallel component of electron Fermi
wavevector, quasiparticle FSAD pattern developed at the image plane is
zeroth-order minimum if $k_x x=n\pi$ ($n$ is an integer and $x$ is SC layer
thickness). In contrast, if $\Delta(-k_x,\mathbf{k}_\parallel)=-\Delta(k_x,%
\mathbf{k}_\parallel)$, the corresponding FSAD pattern is zeroth-order
maximum. Observable consequences are discussed for iron-based
superconductors of complex multi-band pairings.
\end{abstract}

\maketitle

\affiliation{$^1$Department of Physics, National Taiwan Normal University, Taipei 11677,
Taiwan\\
$^2$Department of Physics, Yanshan University, Qinhuangdao 066004, China}


Recently high-$T_c$ superconductivity has been observed in several classes
of Fe-pnictide materials 
\cite{kamihara, Hsu}. 
One key issue for understanding the superconductivity of pnictides lies on
the pairing symmetry of the Cooper pairs. A conclusive observation of the
pairing symmetry remains unsettled to which both nodal and nodeless order
parameters were reported in recent experiments, however. ARPES measurements
indicated clearly a nodeless gap at all points on Fermi surfaces (FS) \cite%
{Ding_EPL,zhao} and magnetic penetration depth measurements further
suggested the gap being possibly in the $s^{\pm}$ state
~\cite{PhysRevLett.102.127004, Martin, mazin:057003}.
The $s^\pm$ state is currently a promising pairing candidate for iron pnictides,
which changes sign between $\alpha$ and $\beta$ bands
and can be naturally explained by the spin fluctuation mechanism
\cite{mazin:057003, wang:047005, Tsuei2010}.
On the contrary, the scanning SQUID microscopy measurements seemed to
exclude the spin-triplet pairing states and suggested the order parameter
having well-developed nodes [\onlinecite{hicks-2008}]. In addition, NMR
experiments were also in favor of nodal SC order
parameters \cite{Matano,Grafe08}. For phase sensitive experiments,
one point-contact spectroscopy
reported was in favor of a nodal gap \cite{Shan}, while the other reported
was in favor of a nodeless gap \cite{chen}.



The complex pairing symmetry of these materials suggests that the pairing
mechanism is likely non-universal and may depend strongly on the fine
details of the band structures. With this regard, some possible experiments
were proposed to elucidate these issues \cite{zhou09, Feng2009,
Lin2010}. In this paper, Fresnel single aperture diffraction (FSAD) of
electrons is proposed as a phase-sensitive probe for studying the SC pairing
symmetry. Of particular interest, it is suggested that FSAD could be very
useful for studying the iron-pnictide superconductors of complex multiple FS
pairings. It will be shown that large $Z$ (effective potential barrier)
zeroth-order FSAD pattern is sensitive to both the SC phase and the probing
direction and thus can give an unambiguous signal to distinguish different
pairing symmetries among different FSs.





\begin{figure}[ptb]
\vspace{-1.5cm}
\par
\begin{center}
\includegraphics[width=1.4\columnwidth]{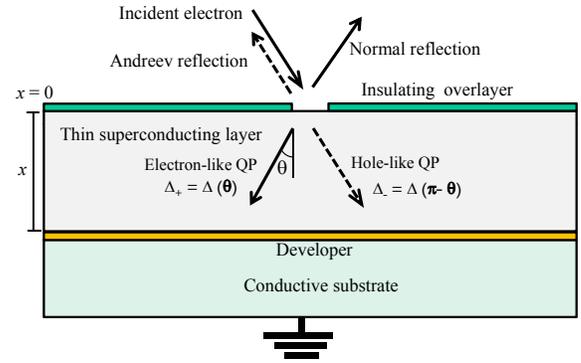}
\end{center}
\par
\vspace{-11.0cm}
\caption{(Color online) Schematic illustration of the Fresnel single
aperture diffraction experiment for a thin superconductor layer (with
thickness $x$). Reflection and transmission processes of a NIS tunnelling
junction are shown. The developer is where the diffraction pattern is
recorded.}
\label{fig1}
\end{figure}

Fig.~\ref{fig1} sketches the proposed apparatus of a FSAD experiment. A
substrate layer, made of a good conductor, is grown firstly. Next, a sheet
of electron-density sensitive developer for recording the diffraction
pattern is deposited. The developer can be made either by the
electron-sensitive material (like the photographic film) or alternatively by
the fluorescent material (like the TV screen).
On top of the recording sheet, a thin layer of SC single crystal with
desired orientation and thickness is assembled. The last step is to coat an insulating
layer on the thin SC layer with a small circular aperture (by mechanical
and/or optical method) for electron beam injection. While electron beam can
be made by natural radioactivity or low-energy accelerator, it can be
alternatively due to a sharp conductive STM tip by an applied voltage bias.
For the latter case, an insulating layer is not needed because the
separation between the tip and the SC thin layer already acts like an
insulating layer between it.

When electrons tunnel into the superconductor through the circular aperture,
matter waves can interfere constructively or destructively. With enough
electrons passing through, clear diffraction pattern can be recorded in the
developer while extra electrons will flow into the ground (see Fig.~\ref{fig1}).
To maintain the coherence for the FSAD signal, the thickness of the thin SC
layer should be made comparable to the SC coherence length. Moreover,
quasiparticles needs to be in the ballistic transport regime or the signal
will be less clear. Scanning
tunneling spectroscopy and vortex imaging have revealed that iron-pnictide
superconductors have a short coherence length, $\xi \approx 27.6\pm 2.9$\AA %
~ \cite{Yin:097002}, comparable to that of cuprate superconductors ($\xi
\leq 20$\AA ) \cite{Pan}. Nevertheless, modern film growing technique has
recently advanced that by improving the quality of the substrate which
minimizes the inverse proximity effect, a nearly perfect ultrathin high-$%
T_{c}$ SC layer can be grown as thin as three unit cells only \cite{Logvenov}%
. This makes the proposed FSAD experiment feasible.

Quasiparticle (QP) states of an inhomogeneous superconductor have a coupled
electron-hole character and can be described by the BdG equations \cite{Gennes}
\begin{eqnarray}
E_{\mathbf{k}}u &=&h_{0}u+\Delta _{\mathbf{k}}v  \notag \\
E_{\mathbf{k}}v &=&\Delta^*_{\mathbf{k}}u-h_{0}v,
\label{Bogoliubov-de Gennes equations}
\end{eqnarray}
where $h_{0}\equiv -\hbar ^{2}\nabla^{2}/2m-\mu$
with $\mu$ the chemical potential and $m$ the electron mass. We consider the
quantum tunneling in an NIS junction where the thin SC layer is made normal
to the $x$-axis and the pairing potential is assumed to be $\sim \Delta _{%
\mathbf{k}}\Theta(x)$ with $\Theta \left( x\right)$ the Heaviside step
function and $\Delta _{\mathbf{k}}$ the SC gap function in the momentum
space \cite{Hu1526}.
For simplicity, proximity effect of the SC order parameter
is ignored at the interface. Under the WKBJ approximation \cite{Bardeen556},
QP wavefunctions for the SC thin layer side ($x>0$) are
\begin{equation}
\left(
\begin{array}{c}
u \\
v%
\end{array}%
\right) =\left(
\begin{array}{c}
{e}^{i\mathbf{k}_{F}\cdot \mathbf{r}}\tilde{u} \\
{e}^{-i\mathbf{k}_{F}\cdot \mathbf{r}}\tilde{v}%
\end{array}%
\right) ~\mathrm{and}~\left(
\begin{array}{c}
\tilde{u} \\
\tilde{v}%
\end{array}%
\right) =e^{-\gamma x}\left(
\begin{array}{c}
\hat{u} \\
\hat{v}%
\end{array}%
\right),  \label{eq:WKBJ}
\end{equation}%
where $\mathbf{k}_{F}\equiv(k_x,\mathbf{k}_\parallel)$ is the Fermi
wavevector and $\gamma $ is the attenuation constant. 
With Eq.~(\ref{eq:WKBJ}), Eq.~(\ref{Bogoliubov-de Gennes equations}) can be
reduced to the Andreev equation
\begin{equation}
E_{\mathbf{k}}\left(
\begin{array}{c}
\hat{u} \\
\hat{v}%
\end{array}%
\right) =\left(
\begin{array}{cc}
\varepsilon & \Delta _{\mathbf{k}} \\
\Delta _{\mathbf{k}} & -\varepsilon%
\end{array}%
\right) \left(
\begin{array}{c}
\hat{u} \\
\hat{v}%
\end{array}%
\right),  \label{Andereev equations for SC}
\end{equation}%
where $\varepsilon \equiv i\gamma k_{x}/m$.
The wavevector parallel to the interface, $\mathbf{k}_\parallel$, is
conserved during the processes \cite{Tanaka3451}.

Solving Eq.~(\ref{Andereev equations for SC}), one obtains two degenerate
eigenstates corresponding respectively to electron- and hole-like QPs:
\begin{equation}
\psi _\mathbf{k}^e(\mathbf{r})=\left(
\begin{array}{c}
\Delta _{+} \\
E_{\mathbf{k}}-\varepsilon
\end{array}%
\right){e}^{i\mathbf{k}_{F}\cdot \mathbf{r}};~
\psi _\mathbf{k}^h(\mathbf{r})=\left(
\begin{array}{c}
E_{\mathbf{k}}+\varepsilon  \\
\Delta _{-}%
\end{array}%
\right){e}^{-i\mathbf{k}_{F}\cdot \mathbf{r}},
\end{equation}%
where $E_{\mathbf{k}}=\sqrt{\Delta _{\mathbf{+}}^{2}+\varepsilon ^{2}}$,
$\Delta _{+}\equiv \Delta (k_{x},\mathbf{k}_{\parallel })=\Delta \left(
\theta \right)$, and  $\Delta _{-}\equiv \Delta (-k_{x},\mathbf{k}_{\parallel })=\Delta
\left( \pi -\theta \right) $  (scattering angle $\theta $ is defined in Fig.~\ref{fig1}).
Superposition of the above two eigenstates
will give a resulting wave function for the SC layer
\begin{equation}
\psi _{S}(\mathbf{r})=e^{-\gamma x}\left[ c_{1}\psi^e_{\mathbf{k}}\left(
\mathbf{r}\right) +c_{2}\psi^h_{\mathbf{k}}\left( \mathbf{r}\right) %
\right].  \label{eq:psiS}
\end{equation}%
The coefficients $c_1$ and $c_2$ are important which are to
be determined by the boundary conditions.
Apart from the factor $e^{-\gamma x}$, Eqs.~(\ref{eq:WKBJ})--(\ref{eq:psiS})
give explicitly
\begin{equation}
\begin{array}{c}
\hat{u}\left( \mathbf{r}\right) =c_{1}\Delta _{+}{e}^{i\mathbf{k}_{F}\cdot \mathbf{%
r}}+c_{2}\left( E_{\mathbf{k}}+\varepsilon \right) {e}^{-i\mathbf{k}%
_{F}\cdot \mathbf{r}}, \\
\hat{v}\left( \mathbf{r}\right) =c_{1}\left( E_{\mathbf{k}}-\varepsilon \right) {e}%
^{i\mathbf{k}_{F}\cdot \mathbf{r}}+c_{2}\Delta _{-}{e}^{-i\mathbf{k}%
_{F}\cdot \mathbf{r}}.%
\end{array}
\label{wave function}
\end{equation}

When an electron is injected into the SC layer through the aperture, there
are two types of reflections: normal reflection of electrons (with the
coefficient $r_N$) and Andreev reflection of holes (with the coefficient $%
r_A $). In terms of $r_N$ and $r_A$, the resulting wave function for the
normal side ($x<0$) can be written as
\begin{equation}
\psi _{N}\left( \mathbf{r}\right) =\left[
\begin{array}{c}
{e}^{i\mathbf{k}_{F}\cdot \mathbf{r}}+r_N{e}^{-i\mathbf{k}_{F}\cdot \mathbf{r%
}} \\
r_A{e}^{i\mathbf{k}_{F}\cdot \mathbf{r}}%
\end{array}%
\right].  \label{eq:psiN}
\end{equation}%
By applying the following boundary conditions:
\begin{eqnarray}
&&\psi _{N}\left( \mathbf{r}\right) |_{x=0^{-}} =\psi _{S}\left( \mathbf{r}%
\right) |_{x=0^{+}},  \label{eq:boundary conditions} \\
&& \frac{2mH}{\hbar ^{2}}\psi _{S}\left( \mathbf{r}\right) |_{x=0^{+}}=\frac{%
d\psi _{S}\left( \mathbf{r}\right) }{dz}|_{x=0^{+}}-\frac{d\psi _{N}\left(%
\mathbf{r}\right) }{dz}|_{x=0^{-}}  \notag
\end{eqnarray}%
with $H$ the height of the delta-function potential for the
barrier, coefficients in (\ref{wave function}) are
solved to be
$c_1=\Delta _{-}(1-iZ)/D$ and $c_2=iZ(E_{\bf k}+\varepsilon)/D$ with
$D=\Delta _{+}\Delta_{-}(1+Z^{2})-Z^{2}(E_{\bf k}-\varepsilon)^2$
and $Z\equiv 2mH/\hbar^{2}k_{F}$ being the effective potential barrier.

In general, the diffraction pattern recorded in the developer
will be proportional to the QP density
developed on it. In the current case, the FSAD intensity is proportional to
\begin{equation}
I({\bf r})=\frac{1}{S}\sum_{i,\mathbf{k}}\left[\left\vert
\hat{u}_i\left(\mathbf{r}\right) \right\vert ^{2}f\left(
E_{\mathbf{k}}\right) +\left\vert
\hat{v}_i\left(\mathbf{r}\right) \right\vert ^{2}f\left( -E_{\mathbf{k}}\right) %
\right] \Delta S_{i},  \label{electron number distribution}
\end{equation}%
where $f(E_{\mathbf{k}})=(e^{\beta E_{\mathbf{k}}}+1)^{-1}$
and considering the
size effect, a spatial average over the aperture
(of area $S$) is taken where $\Delta S_{i}$ denotes a tiny cell within $S$.


\emph{Gap symmetry and barrier $Z$ dependent FSAD} -- For simplicity, we
shall consider the limit such that aperture diameter $d$ is much smaller
than the thickness $x$ of the SC layer. This means that the spatial average
in (\ref{electron number distribution}) is not needed. In the limit of $%
T\rightarrow 0$, interesting results of SC gap symmetry and barrier $Z$
dependent FSAD will be obtained.
Knowing the coefficients $c_1$ and $c_2$,
QP wavefunctions developed at ${\bf r}=(x,0,0)$ on the developer are obtained
to be
\begin{eqnarray}
\hat{u}(x) &=& \hat{v}(x)  \label{eq:sp} \\
&=& \left\{
\begin{array}{ll}
e^{ik_{x}x}-2iZ\sin(k_{x}x), & ~\mathrm{if}~\Delta_{-}=\Delta_{+} \\
e^{ik_{x}x}-2iZ\cos(k_{x}x), & ~\mathrm{if}~\Delta_{-}=-\Delta_{+}~.%
\end{array}
\right.  \notag
\end{eqnarray}
There are many observable consequences
out of the above symmetry-dependent results. In the following, we illuminate
how one can probe the pairing symmetry and Fermi surface of a superconductor
based on Eq.~(\ref{eq:sp}).

When barrier is low, $Z\ll 1$, $\hat{u}(x)= \hat{v}(x)=\exp \left( ik_{x}x\right)$ for
both even ($\Delta _{-}=\Delta _{+}$) and odd ($\Delta _{-}=-\Delta _{+}$)
symmetries. In this limit, FSAD pattern recorded in the developer makes no
difference between the two symmetries.
In this case, $I=1$ and a zeroth-order maximum FSAD pattern (Airy disk)
will occur. 
Nevertheless, the $Z\ll 1$ FSAD pattern can be used to measure the FS of the
SC sample. 
Using the well-known formula $\sin \theta=1.22\lambda/d$ ($d$ is the
aperture diameter) that locates first minimum of the Airy pattern, one can
measure $\theta$ which determines the de Broglie wavelength of electrons, $%
\lambda$, and hence unambiguously identify the Fermi vector along the
particular direction via the relation, $k_x=2\pi/\lambda$. It should be
emphasized that the above result remains valid even when $Z$ is not too
small ($Z\alt 1$).

\begin{figure}[ptb]
\vspace{-0.5cm} 
\includegraphics[width=1.15\columnwidth]{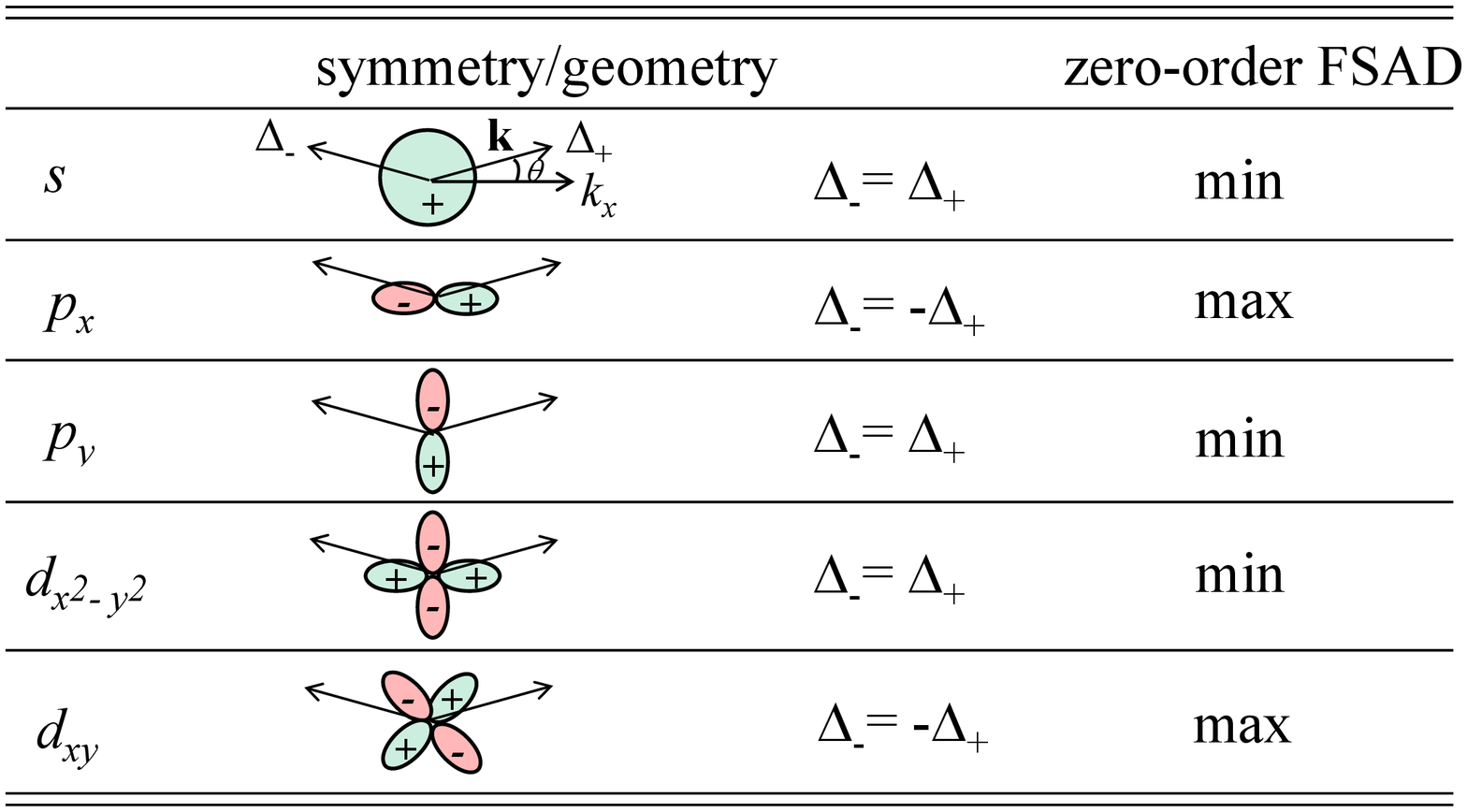} 
\vspace{-3.0cm}
\caption{(Color online) Illustration of SC gap symmetry dependent FSAD in
the large-$Z$ limit.}
\label{fig2}
\end{figure}

Of most interest is when the barrier is high, $Z\gg 1$, to which \emph{only the
first Fresnel zone appears when $d$ is small enough}. In this limit, $\hat{u}(x)=
\hat{v}(x)\sim\sin (k_x x)$ for even symmetry and $\sim\cos (k_x x)$ for odd symmetry.
Consequently
\begin{equation}
I\sim\left\{
\begin{array}{ll}
\sin^2 (k_x x), & ~~~~\mathrm{if}~\Delta _{-}=\Delta _{+} \\
\cos^2 (k_x x), & ~~~~\mathrm{if}~\Delta _{-}=-\Delta _{+}%
\end{array}%
\right.  \label{nn}
\end{equation}%
and the zeroth-order FSAD pattern developed behaves drastically different
between the two symmetries.
Experimentally to create a high barrier $Z$ a thin insulating layer can be
coated on the SC layer in assembling the FSAD apparatus. Alternatively, $Z$
can be tuned by adding a bias voltage in the substrate layer relative to the
ground, in addition to the bias voltage between the tip and the substrate
layer. 
Moreover, for the large-$Z$ limit, it is well-known that for even-parity
pairing, maximum conductance occurs when incident electron energy equals the
gap amplitude, $E\approx \Delta$. While for odd-parity pairing, owing to the
zero-bias bound (midgap) state, maximum conductance occurs at $E\approx 0$
\cite{Tanaka3451}. Thus for the present FSAD experiment, one can try to tune
the incident electron energy to gain maximum-intensity signal.

Knowing the Fermi vector $k_x$ (at particular direction), one may grow the
SC film for the FSAD experiment with the desired thickness $x$
which satisfies $k_x x=n\pi$ ($n$ is an integer) and is comparable to the
coherence length $\xi$. Consequently for even symmetry, $I\sim\sin^2 \pi$ and
FSAD will show a zeroth-order \emph{minimum} at the developer. In contrast,
for odd symmetry, $I\sim\cos^2 \pi$ and the FSAD will show a zeroth-order \emph{%
maximum}. 

Fig.~\ref{fig2} illustrates the large-$Z$ gap-symmetry dependent FSAD
pattern for different symmetries.
While iron-pinicides seem to be spin-singlet superconductors with possibly $s$-
and/or $d$-wave pairing symmetries [\onlinecite{hicks-2008}], for completeness
and for references to a spin-triplet superconductor of interest, we have
also considered the cases of $p$-wave symmetry in Fig.~\ref{fig2} (and also
in Table~\ref{table1}).
As is shown, for all cases we consider that the
direction of injected electron, $\mathbf{k}$, is pointing near the $k_x$
axis (with an angle $\theta$). For $s$-wave gap, $\Delta _{-}$= $\Delta_{+}$
and the corresponding FSAD will be a zeroth-order minimum, which is
apparently independent of the direction of electron injected. However, for
nodal superconductors such as $p$- and $d$-wave cases, the results are two
folds. If $\mathbf{k}$ is pointing such that $\Delta _{-}$= $\Delta_{+}$
(for example the $p_y$ and $d_{x^2-y^2}$ symmetries in Fig.~\ref{fig2}), the
corresponding FSAD will show a zeroth-order minimum. In contrast, if $%
\mathbf{k}$ is pointing such that $\Delta_{-}$= $-\Delta_{+}$ (for example
the $p_x$ and $d_{xy}$ symmetries in Fig.~\ref{fig2}), the corresponding
FSAD will show a zeroth-order maximum. It is worth noting that for all $p$-
and $d$-wave nodal cases, if $\mathbf{k}$ is pointing right at the nodes, $%
\Delta _{+}=\Delta_{-}=0$, the corresponding FSAD pattern will show a
zeroth-order maximum, analogous to the case of a normal metal.

\begin{figure}[ptb]
\vspace{-1.5cm}
\par
\begin{center}
\includegraphics[width=0.8\columnwidth]{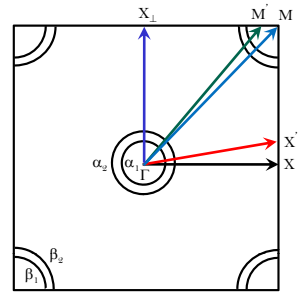}
\end{center}
\par
\vspace{-5.4cm}
\caption{(Color online) Schematic of the Fermi surfaces of Fe-pnictide
superconductors in folded Brillouin zone. Important incident electron
directions for FSAD experiment are shown.}
\label{fig3}
\end{figure}

Moreover, for both $d_{x^2-y^2}$ and $d_{xy}$ symmetries, zeroth-order FSAD pattern
could change from maximum (minimum) to minimum (maximum) if the SC layer is
grown with $\pi/4$ rotated about the $c$-axis (assuming that SC gap mainly
develops in the $ab$ plane). Similarly, for both $p_x$ and $p_y$ symmetries,
zeroth-order FSAD pattern could change from maximum (minimum) to minimum
(maximum) if the SC layer is grown with $\pi/2$ rotated about the $c$-axis.
This gives another machinery for FSAD to distinguish between $s$-, $p$-, and $d$-wave
pairing symmetries.


We now discuss possible schemes of FSAD patterns for multiband iron-pnictide
superconductors. The so-called $\alpha$ sheets are concentric and nearly
circular hole pockets around the $\Gamma$ point. While the $\beta$ sheets
are nearly circular electron pockets around the M points \cite{singh:237003,
cao:220506}. These FS sheets are sketched in Fig.~\ref{fig3}. If the pairing
originates from the same mechanism, most likely $\alpha_1$ and $\alpha_2$
bands will have the same pairing symmetry. Similarly $\beta_1$ and $\beta_2$
bands will also likely have the same pairing symmetry. However, pairing
symmetries may differ between $\alpha$ and $\beta$ bands. 
Among other experiments,
one can actually perform FSAD experiment to test the pairing symmetry of
each \emph{individual} band by carefully tuning the energy of incident
electrons for maximum intensity with desired orientation and layer
thickness.
\begin{table}[t]
\caption{Possible FSAD patterns for various pairing symmetries and incident
directions shown in Fig.~\protect\ref{fig3}.}
\begin{ruledtabular}\label{table1}
\begin{tabular}{cccccccc}
symmetry & $\Gamma$X & $\Gamma$X$^\prime$ & $\Gamma$M & $\Gamma$M$^\prime$ & $\Gamma$X$_\bot$ \\
\hline
$s$ & min & min & min & min & min\\
$p_x$ & max & max & max & max & max \\
$p_y$ & max & min & min & min & min\\
$d_{x^2-y^2}$ & min & min & max & min & min\\
$d_{xy}$ & max & max & max & max & max\\
\end{tabular}
\end{ruledtabular}
\end{table}

More explicitly, one can first grow a set of thin layers with different
crystal directions, and then perform small-$Z$ FSAD experiments to measure
and compile the FSs. With the knowing FSs, one can grow another set of thin
layers with desired thickness $x$ and crystal directions. For instance, if
one thin layer has $x$ simultaneously satisfying $k_1 x= n_1\pi$ and $k_2 x=
n_2\pi$ with $n_1,n_2$ both integers and $k_1$ and $k_2$ the corresponding
Fermi vectors of $\alpha_1$ and $\alpha_2$ (or $\beta_1$ and $\beta_2$)
bands, one can then perform large-$Z$ FSAD experiment on this thin layer to
sort out the pairing symmetry on $\alpha$ and/or $\beta$ bands. Taking LaO$%
_{1-x}$F${_x}$FeAs as an example, if $k_x\simeq 0.22\pi/a$ for $\beta$-band with
lattice constant $a\simeq 0.4$nm \cite{raghu_prb_08,Takahashi_nature_08},
the thickness of the SC film can be better taken to be $x=n\pi/k_x\simeq n(4.55a)\simeq
n(1.84\mathrm{nm})$.
For $n=2$, $x\simeq 3.68 \mathrm{nm}=36.8$\AA .

In Fig.~\ref{fig3}, important directions of incident electrons of FSAD
experiment are indicated for iron-pnictides. Possible FSAD patterns for
various incident directions and pairing symmetries are listed in Table~\ref
{table1}. Note that it is also important to do the FSAD experiment for
the $\Gamma$X$^\prime$ and $\Gamma$M$^\prime$ directions which are slightly
deviated from the $\Gamma$X and $\Gamma$M directions. In view of Table~\ref{table1},
if the zeroth-order FSAD pattern changes from maximum for $%
\Gamma$X to minimum for $\Gamma$X$^\prime$ direction, pairing symmetry is
likely to be $p_y$-wave. Similarly, it is likely to be $d_{x^2-y^2}$-wave if
it changes from maximum for $\Gamma$M to minimum for $\Gamma$M$^\prime$
direction. 

The proposed FSAD experiment is sensitive to the pairing gap symmetry on one
particular FS. Due to the nature of a zero momentum transfer
probe, it cannot link the pairing gap symmetries on two distant FSs.
For iron-pnictides, one can use the experiment to check whether
it's $s$-wave on both $\alpha$ and $\beta$ bands which is consistent with the $s_\pm$
state, or $s$-wave on one band and $d$-wave on the other band. However, it
is not able to tell if there is a sign change between the two bands. To
verify the sign change, other experiments which can link the pairing
symmetries on two distant FSs are in demand.

In summary, we propose that Fresnel single aperture diffraction (FSAD) could
be a useful phase-sensitive probe for the pairing symmetry of a
superconductor. It is demonstrated that FSAD pattern is intimately related
to the SC pairing symmetry and the direction of incident electrons.
Possible designs of FSAD experiment are suggested and discussed for
iron-pnictide superconductors of complex multiple Fermi surface pairings. It
is noted that the same scheme discussed in the present paper can also
be applied to other phase-sensitive experiments, such as Young's
interference and Fresnel lens.

\begin{acknowledgments}
This work was supported by National Science Council of Taiwan (Grant No.
99-2112-M-003-006),
Hebei Provincial Natural Science Foundation of China (Grant
No. A2010001116),
and National Natural Science Foundation of China (Grant No. 10974169).
We also acknowledge the
support from the National Center for Theoretical Sciences, Taiwan.
\end{acknowledgments}

%


\end{document}